%% file: mm.tex
\documentclass{acm_proc_article-sp}

\pdfoutput=1
\usepackage{latexsym}
\usepackage{booktabs}
\usepackage{amsmath,amssymb}
\usepackage[boxed]{algorithm}
\usepackage{algorithmic}
\usepackage{graphicx}
\usepackage{multirow}
\usepackage[tight]{subfigure}

\usepackage[pdftex,naturalnames,pdfpagemode={UseOutlines},letterpaper,bookmarks,bookmarksopen=true,pdfstartview={FitH},bookmarksnumbered=true,pdftitle={-},pdfauthor={-}]{hyperref}

\renewcommand{\paragraph}[1]{\vspace{1mm}\noindent{\bf #1}}
\newcommand{\eat}[1]{}

\newtheorem{definition}{Definition}[section]

\newtheorem{theorem}[definition]{Theorem}

\newtheorem{example}[definition]{Example}

\newcommand{\squishlist}{
  \begin{list}{$\bullet$}
   {
     \setlength{\itemsep}{0pt}
     \setlength{\parsep}{3pt}
     \setlength{\topsep}{3pt}
     \setlength{\partopsep}{0pt}
     \setlength{\leftmargin}{1.5em}
     \setlength{\labelwidth}{1em}
     \setlength{\labelsep}{0.5em} } }

\begin{document}

\title{Matrix Multiplication Using Only Addition}
\numberofauthors{2}
\author{Daniel L. Cussen$^{\dag}$, Jeffrey D. Ullman$^{\sharp}$\\
\affaddr{\large $^{\dag}$Fgemm SPA,
$^{\sharp}$Stanford University}\\
}

\maketitle

\input{mm1}

\end{document}

%% file: mm1.tex
\begin{abstract}
Matrix multiplication consumes a large fraction of the time taken in many machine-learning algorithms.  Thus, accelerator chips that perform matrix multiplication faster than conventional processors or even GPU's are of increasing interest.  In this paper, we demonstrate a method of performing matrix multiplication without a scalar multiplier circuit.  In many cases of practical interest, only a single addition and a single on-chip copy operation are needed to replace a multiplication.  It thus becomes possible to design a matrix-multiplier chip that, because it does not need time-, space- and energy-consuming multiplier circuits, can hold many more processors, and thus provide a net speedup.
\end{abstract}

\section{Introduction}
\label{intro-sect}

In this paper we show that when multiplying matrices, scalar multiplication is not really necessary and can be replaced by a surprisingly small number of additions.  The advantage of performing matrix multiplication using only addition for arithmetic is that it then becomes feasible to build special-purpose chips with no multiplier circuits.  Such chips will take up less space per on-chip processor, allowing more, but simpler, processors to be packed into a single chip.  \cite{plasticine} is an example of an architecture that implements this strategy.  Further, since a multiplier circuit can take significantly more time than addition or other typical machine operations, it is possible that the addition-only approach can be faster, even on conventional architectures.  But there is little advantage if the number of additions needed to replace a single multiplication is large.  Thus, we must show that in practice, very few additions are needed to replace a multiplication~-- fewer than one addition (plus a little more than one copy operation) in some cases of practical interest.

Before getting deep into the weeds, let us see the essential idea that the algorithm uses.  Start with a vector of integers in some large range, say 1-to-$k$.  We sort, eliminate duplicates, and take differences between consecutive numbers in the sorted order.  The result is a new vector of integers, but they are much more constrained than the original list.  Specifically, rather than simply being numbers in the range up to $k$, their {\em sum} is at most $k$.  Thus, when we iterate the process of sorting, eliminating duplicates, and taking differences several times, the lists rapidly become much shorter than the original list.  We can then multiply the original vector by any constant $c$ by recursively multiplying the vector of differences by $c$ and then obtaining the original vector by accumulating the differences after they have been multiplied by $c$.

\subsection{Motivation and Background}
\label{refs-subsect}

The rising importance of deep-learning and other machine-learning applications has made multiplication of large matrices take on a new importance.  For example, backpropagation \cite{backprop} is essentially a sequence of matrix multiplications.  At the same time, special-purpose chips or boards, such as \cite{plasticine} \cite{TPU}, are proliferating.  We therefore offer a new approach to matrix multiplication that:

\begin{enumerate}

\item
Works for both sparse and dense matrices.

\item
Is more efficient, the larger the matrices are.

\item
Works better when the matrix elements require fewer bits, an important trend as people search for ways to make machine learning more efficient.

\item
Supports a chip design that avoids multiplication circuits, thus saving chip real estate and allowing more arithmetic units to be placed on one chip.

\item
Uses a very small number of additions in place of one scalar multiplication, thus offering an opportunity to  speed up the computation, since multiplication can take significantly more time than addition.

\end{enumerate}

The search for algorithms that multiply $n$-by-$n$ matrices in less than the $O(n^3)$ time taken by the straightforward algorithm has been ongoing for more than 50 years, from Strassen \cite{strassen} at $O(n^{2.81})$ to the best known \cite{williams} at $O(n^{2.37})$.  Unfortunately, all these algorithms to date, while they have better asymptotic running times than the obvious, have constant factors that make them unattractive, even for very large matrices, and they also assume the matrices are dense.

Our central thesis is that it makes sense to replace multiplication by a small number of additions.  This point can only be justified if the time taken for multiplication significantly exceeds the time taken for addition.  \cite{fog} is a recent examination of the time taken by all instructions on many of the most commonly used processors.  The conclusion is that typically, integer multiplication has 3–6 times the latency of integer addition.\footnote{When we talk of floating-point operations, the difference is much less.  The reason is that when multiplying, we have only to add exponents, while a floating-point addition requires that we align the mantissas according to the difference in the exponents.  However, we are not proposing to replace floating-point multiplications by floating-point additions.  We are proposing to replace the multiplication of mantissas, which are integers, by integer addition.}  Thus, replacing multiplication by one or two additions looks promising.

The issue of how much space and energy a multiplier takes on a chip also argues that there are potential advantages to avoiding using a multiplier altogether.  The faster multipliers involve complex circuits ({\em compressors}, where several of the partial products are combined not into a single sum, but into two or more integers that themselves must be combined further \cite{wikimult}.  Even more space- and energy-consuming circuits allow {\em pipelined} multiplication, where the same circuit can be used to do several multiplications at once.  Since it is then possible, in principle,  to produce the result of one multiplication in the same time as it takes to perform one integer addition, the utility of our addition-only method depends on being able to fit several adders in the space that one pipelined multiplier takes, and/or use less energy with the adder-only approach.

There is another approach to addition-only matrix multiplication.  One can use a table lookup to convert numbers to the log domain, add the logarithms, and convert back \cite{logs}.  In this way, one multiplication is replaced by one addition.  This method works, although its efficiency depends on how large the log and antilog tables are, which in turn depends on how precise we wish the result to be.  And at any rate, the cost of the two lookups must be included in the total cost.

\subsection{Simplifying Assumptions}
\label{simp-subsect}

To describe simply the ideas behind our addition-based approach to matrix multiplication, we shall make several assumptions.  First, we assume that we are multiplying $n$-by-$n$ matrices, and that the elements of these matrices are $b$-bit integers.  Typically, $b$ would be 32 or 24, but even smaller integers are becoming more interesting.  For example, the TPU (tensor-processing unit) \cite{TPU} is based on a 16-bit floating-point number, which has effectively a 12-bit mantissa, counting the implied leading 1.  Note that if elements are single-precision floats, they have (effectively) a 24-bit mantissa, and it is only the mantissas that need to be multiplied.  The exponent portions are added, and there may be a shift of a single position needed in the product of mantissas.  Section~\ref{extensions-subsect} will show how the ideas presented here can be extended more broadly, including to matrices of arbitrary shapes and to sparse matrices.

We assume that our processors have a ``shift-and-add'' operation $\mathrm{add}(x,y,i)$.  This operation shifts the integer $y$ left by $i$ positions and adds the result to $x$.  We assume $y$ is a $b$-bit integer, $x$ can be as large as $2b-1$ bits, and $0\le i<b$.  That is, the effect of this operation is
$$x~:=~x + 2^iy$$
In addition, we shall assume processors have available the operations needed to sort, eliminate duplicates, connect elements of an initial list to their representation in the sorted list (i.e., follow pointers), and take the first differences of a sorted list.  However, since the total number of operations of these types needed in the proposed algorithm is $O(n^2\log n)$, and thus small compared with the total number of operations, we shall not count the operations of these types exactly.

\subsection{Russian-Peasants Multiplication}
\label{rpm-subsect}

Suppose we multiply $n$-by-$n$ matrices $P$ and $Q$ with elements $[p_{ik}]$ and $[q_{kj}]$, respectively,  each of which is a $b$-bit integer.  the product has in position $(i,j)$ the value $\sum_{k=1}^n p_{ik}\times q_{kj}$.  We therefore have to take $n^3$ products of $b$-bit integers.  Another $(n-1)n^2$ additions are needed to sum the terms that form each of the $n^2$ result elements, but we need those additions no matter how the multiplications are done.  Thus, in discussions that follow, we shall not count these additions.

As a baseline for replacing multiplication by addition, we can always simulate a multiplication as a sequence of shift-and-add operations.  This technique, which is often referred to as ``Russian-Peasants Multiplication,'' uses only addition and was apparently known as far back as ancient Egypt \cite{rpm}. The method is usually explained in decimal, but when numbers are expressed in binary, it is especially simple.  Suppose we are multiplying $p\times q$.  Think of $p$ as a bit string $p_{b-1}p_{b-2}\cdots p_0$.  Starting with $x=0$, for each $p_i$ that is 1 we perform the shift-and-add operation $\mathrm{add}(x,q,i)$; the result will be $p\times q$.

We can thus replace $n^3$ multiplications by at most $bn^3$ additions.  If elements of the matrices are chosen at random, then we would expect about half the bits to be 0, and therefore $bn^3/2$ is a better approximation to the number of required additions to simulate $n^3$ multiplications using the Russian-Peasants approach..  However, in fact we can, in practical cases, do much better than $b$ or $b/2$ additions per multiplication.

\section{The Addition-Only Matrix-Mul\-ti\-plic\-a\-tion Algorithm}
\label{algo-sect}

We shall begin by describing the overall approach to matrix multiplication, followed by the details of the basic zero-multiplication algorithm.  In Section~\ref{alignment-subsect}, we shall give an additional idea that can reduce the number of additions needed still further.  Section~\ref{upper-sect} will prove that the algorithm offered does use significantly fewer additions than the Russian-Peasants approach in cases of practical interest.

\subsection{Overview of Matrix Multiplication}
\label{overview-subsect}

The goal is to multiply $n$-by-$n$ matrices $A$ and $B$.  The elements of these matrices are $b$-bit integers.  A common approach is to take $n$ outer products, each is the outer product of a column of $A$ and the corresponding row of $B$.  That is, the $k$th outer product applies to column $[a_{1k},\ldots,a_{nk}]$ of $A$ and the row $[b_{k1},\ldots,b_{kn}]$ of $B$.  The result of this outer product is $n^2$ scalar products $a_{ik}b_{kj}$ for all $i$ and $j$; we add this product to the element of the result that is in row $i$ and column $j$.

The additions that accumulate these products are part of any matrix-multiplication algorithm.  We therefore do not count them when tallying additions made by the proposed method.  That is, we are only counting the additions that we use to replace the $n^3$ scalar products that are needed to perform the $n$ outer products just described.

If we were only to multiply two $b$-bit scalars, we could make some small improvements to the Russian-Peasants approach in some cases, but the latter method is about as good as we can do.  However, when matrices are large and the number of bits in their elements is small ~-- exactly the common case for machine-learning applications~-- we can do significantly better by multiplying a vector of $n$ integers, each of $b$ bits, by a $b$-bit constant.  We preprocess the vector in a way we shall describe, involving sorting, eliminating duplicates, and taking differences of successive numbers.  Once preprocessed, we can multiply the vector by $n$ constants, and thus compute one of the $n$ outer products we need.

\subsection{Vector-Scalar Multiplication}
\label{vector-scalar-subsect}

\begin{figure}[ht]
\centering
\includegraphics[width=0.55\textwidth]{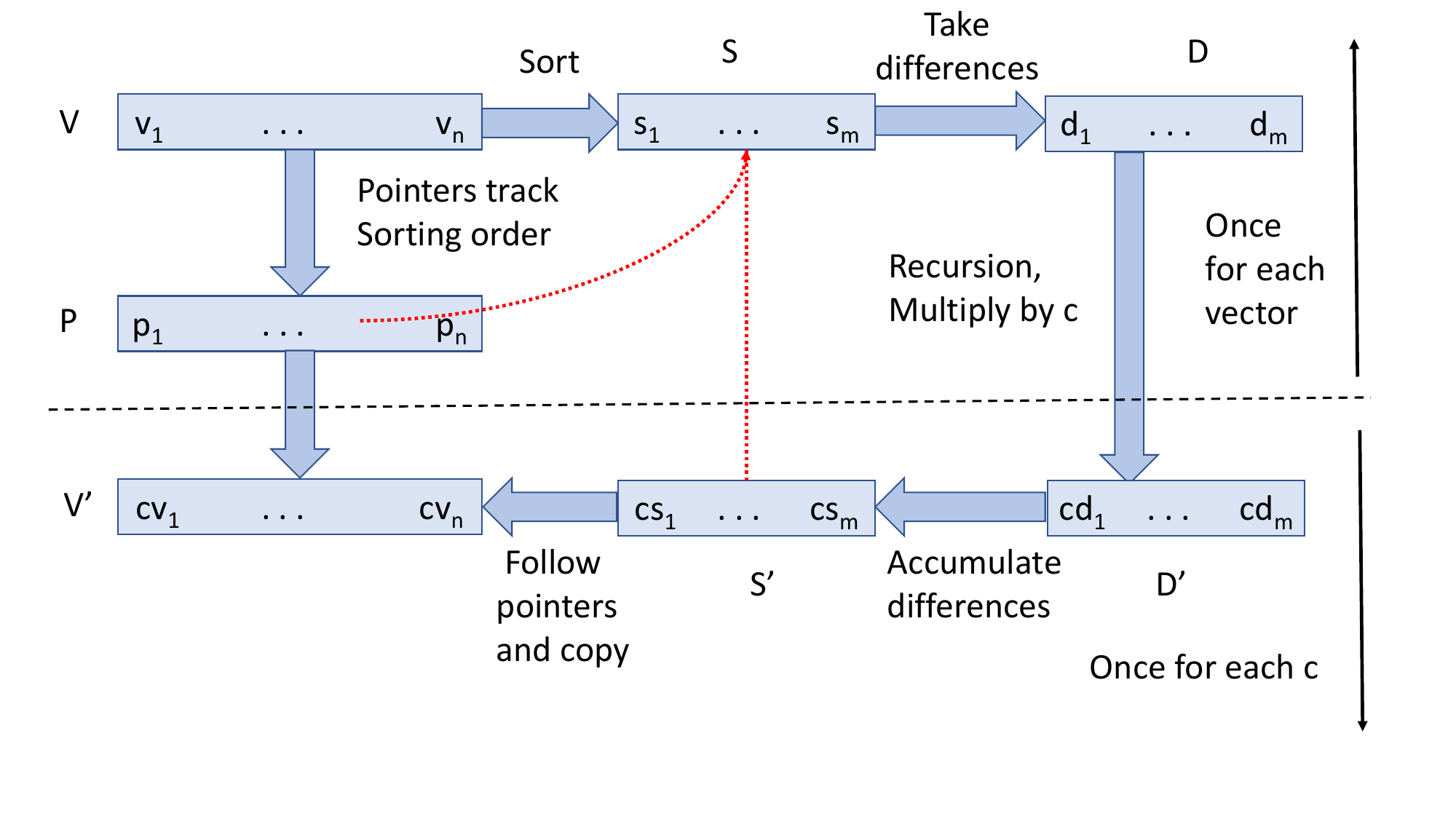}
\caption{Algorithm Outline}
\label{algoutline-fig}
\end{figure}

The algorithm for multiplying a vector $[v_1,\ldots,v_n]$ by a scalar $c$ is suggested by Fig.~\ref{algoutline-fig}.  We assume that each $v_i$ is a positive $b$-bit integer; signed integers can be handled by ignoring the sign until the products are created.  The algorithm is recursive, in that it modifies the initial vector to produce a (typically) shorter vector $[d_1,\ldots,d_m]$.   Not only is this vector shorter, but the sum of its elements is at most $2^b$.  This constraint is much stronger than the constraint on the original vector~-- that each element be less than $2^b$.

 Either using Russian-Peasants multiplication as a base case, or by applying the same ideas recursively, we produce the vector-scalar product $[cd_1,\ldots,cd_m]$, which is then used to produce the desired output $[cv_1,\ldots,cv_n]$.  The recursive part of the algorithm consists of the following steps.

\begin{enumerate}

\item
{\em Sort}: Begin by sorting the given vector $[v_1,\ldots,v_n]$ in a way that eliminates duplicates.  The resulting sorted vector $[s_1,\ldots,s_m]$ has $s_1<s_2<\cdots<s_m$.  It is necessary to create an array of pointers $[p_1,\ldots,p_n]$, where $p_i$ gives the unique value $j$ such that $v_i=s_j$.  These pointers are suggested by the red, dotted lines in Fig.~\ref{algoutline-fig}.\footnote{Depending on how a chip is organized, it may be more efficient to store the array of pointers inverted.  That is, for each $i$ there would be easy access to all the values of $j$ for which $v_j=s_i$.}

\item
{\em Differences}: Construct a new vector $[d_1,\ldots,d_m]$ giving the differences between successive elements of the sorted array, with an imaginary 0 preceding the first number,  That is, $d_1=s_1$ and $d_i = s_i-s_{i-1}$ for $i=2,3,\ldots, m$.

\item
{\em Recursion}: Either by using Russian-Peasants multiplication as a base case, or using this algorithm recursively, produce the vector-scalar product $[cd_1,\ldots,cd_m]$.

\item
{\em Accumulate}: Compute the product of $c$ and the vector $[s_1,\ldots,s_m]$ by $cs_1=cd_1$ and $cs_i=cs_{i-1}+cd_i$ for $i=2,3,\ldots,m$.  Note that we use $m-1$ additions here, and it is the only place, other than the recursive step, where additions are used.

\item
{\em Follow Pointers}: For each $v_i$, the pointer $p_i$ is that $j$ such that $v_i=s_j$.  Therefore, $cp_i=cs_j$.  We may therefore copy the value $cs_j$ just computed at the previous step and make that be $cp_i$.

\end{enumerate}

Observe that while step~(1) requires $O(n\log n)$ time to sort, and step~(2) requires $O(n)$ time for subtractions, these are done only $n$ times each.  Thus, the total time spent for these two steps is $O(n^2\log n)$ at most.  It is only steps (4) and (5), each of which takes $O(n)$ time but is repeated $n^2$ times, that require $O(n^3)$ work.  In particular, step~(4) is where the $O(n^3)$ additions occur, and step~(5) requires $O(n^3)$ copy operations.

\begin{example}
\label{pi-ex}

Suppose we start with the vector
$$V=[3,1,4,1,5,9]$$
When we sort and eliminate duplicates (Step~1), we get the vector $S=[1,3,4,5,9]$ and the vector $P=[2,1,3,1,4,5]$.  That is, the first element of $P$ is 2 because the first element of $V$, which is 3, is in the second position of $S$.  The second element of $P$ is 1, because the second element of $V$, which is 1, has moved to position 1 in $S$, and so on.

Next, take differences of $S$ (Step~2) to get $D=[1,2,1,1,4]$.  That is, the first element of $D$ is the first element of $S$, i.e., 1.  The second element of $D$ is the difference between the first and second elements of $S$, that is, $3-1=2$, and so on.

For the recursion, Step~3, we treat the vector $D$ as if it were $V$.  We sort and eliminate duplicates, to get $[1,2,4]$.  When we take differences, we have $[1,1,2]$.  Sort and eliminate duplicates again, and we get $[1,2]$, whose difference vector is $[1,1]$.  Now, when we sort and eliminate duplicates we have only the vector $[1]$.  We can multiply this vector by any constant $c$ using Russian-Peasants multiplication.\footnote{In fact, multiplication by 1 is trivial, and we really do not need any operation at all in this special case.  However, in rare cases, the limit of the recursion will be a vector whose length is 1 but whose one element is bigger than 1.  It may also make sense to stop the recursion before the vector length reaches 1.}

Suppose that $c=5$; that is, we want to compute $5V$.  Let us assume that recursively, we have managed to compute $D'=5D$, the vector $[5,10,5,5,20]$.  We compute $S'=5S$, the vector $S'=[5,15,20,25,45]$ by accumulating the elements of $D'$ (Step~4).  That is, the first element of $S'$ is the first element of $D'$, i.e. 5.  The second element of $S'$ is the sum of the first element of $S'$ and the second element of $D'$, or $5+10=15$, and so on.

Lastly, we construct $V'=5V$ by applying the vector of pointers $P=[2,1,3,1,4,5]$ to the vector $S'=[5,15,20,25,45]$ (Step~5).  The first element of $V'$ is the second element of $S'$, or 15.  The second element of $V'$ is the first element of $S'$, or 5, and so on.  The result is $V'=[15,5,20,5,25,45]$.

In practice, we would compute a complete outer product by multiplying $V$ by $n$ different constants, of which 5 was just an example.  We use the same $D$ and $P$ vectors for each of the $n$ constants, so there is no duplication of work.
\end{example}

\subsection{How Many Additions?}
\label{count-subsect}

As mentioned, we shall later prove an upper bound on the number of additions needed by this algorithm; see Section~\ref{upper-sect}.  But here, let us look at the intuition why we would expect relatively few additions to be needed.  Intuitively, it is because when $n$ is large and $b$ is small, there can't be too many different differences, and therefore, as we recurse, the lengths of the vectors involved drop quickly.

\begin{example}
\label{count-ex}
Suppose we have a vector of $n=1000$ randomly chosen 12-bit integers.  When we sort them, we expect to find some duplicates, but not many.  So at the first level of recursion, we would have $m$, the length of the sorted list, close to 1000.  But at the second level, we note that the vector of differences, $[d_1,\ldots,d_m]$, has almost 1000 elements that sum to at most $2^{12}=4096$.  Thus, there cannot be many different values among these, and the value of $m$ at the second level will be quite small compared with 1000.  In particular, the sum of the first 91 integers exceeds 4096, so 90 is an absolute upper bound on the number of second differences.

For the third level of recursion, there may be so few that it makes sense to use Russian Peasants on them, or we may recurse a third time, and get a still smaller list.  But the work at levels after the second will be negligible no matter how we do it.

Thus, we require at most 1000 additions at Step~4 of the first level, and at most 91 additions at Step~4 of the second level.  Subsequent levels require even fewer additions, so the total number of additions required at all levels will be just a little more than 1000.  In comparison, if we used Russian Peasants only, we would expect around 6000 additions.
\end{example}

Of course, the count of additions as suggested by Example~\ref{count-ex}  is for one vector-scalar multiplication.  For the complete multiplication of $n$-by-$n$ matrices, we would have to do $n^2$ such operations, for a total of a little more than $n^3$ additions (assuming again that $n=1000$ and $b=12$).  But those additions replace the $n^3$ multiplications that would be needed, so we are still claiming a little more than one addition replacing one multiplication.

\subsection{Running Time of the Complete Algorithm}
\label{running-time-subsect}

Remember that Figure~\ref{algoutline-fig} represents two different phases.  The first phase, above the line, is done once for each row of the second matrix, i.e., $n$ times.  Thus, even though we are sorting a list of length $n$, which takes $O(n\log n)$ time, the total time spent above the line is $O(n^2\log n)$.  That cost can be neglected compared with the $O(n^3)$ running time of the entire algorithm.  However, if we think in terms of chip design, we do have to include on chip the capability of doing the sorting and setting up the pointer array that is implied by Fig.~\ref{algoutline-fig}.

For the operations below the line in Fig.~\ref{algoutline-fig}, they clearly take $O(n)$ time.  The constant factor includes the number of additions needed to replace one multiplication.  For each of the $n$ rows of the second matrix, we perform the operations below the line $n$ times, so the total time taken is $O(n^3)$ as it should be.

In this analysis, we have assumed a serial execution, which is not realistic or desirable if we are to design a special-purpose chip.  Presumably we would design for parallel execution, for example implementing a parallel sort or processing several vector-scalar multiplications at the same time.  However, $O(n^3)$ will still be the measure of the number of operations the chip will execute.

\section{Improvements to the Basic Algorithm}
\label{improvements-sect}

There are a number of modifications to the algorithm of Section~\ref{algo-sect} that improve one or more aspects.  Here, we shall mention {\em alignment} as a technique to reduce the length of vectors involved.  We can take advantage of zeros, ones, and duplicates in the columns of the first matrix.   We also mention ways to increase the available parallelism.  And we address extensions to nonsquare and sparse matrices.

\subsection{Alignment}
\label{alignment-subsect}

If elements $v$ and $w$ of a vector differ by a factor that is a power of 2, then when we multiply the vector by any constant $c$, the products $cv$ and $cw$ will also have a ratio that is the same power of 2.  Therefore, we can obtain $cw$ from $cv$, or vice-versa, by shifting their binary representations.  We can use this observation to treat $v$ and $w$ as if they were the same, if we make some small modifications to the basic algorithm of Fig.~\ref{algoutline-fig}.

\begin{enumerate}
\item
Before sorting $[v_1,\ldots,v_n]$, we shift each element $v_i$ right until it becomes an odd number (i.e., we drop 0's from the lower-order bits).

\item
In addition to the vector of pointers $[p_1,\ldots,p_n]$, we need another vector $H=[h_1,\ldots,h_n]$, where $h_i$ is the number of positions to the right that we have shifted $v_i$. 

\item
When constructing the result vector $[cv_1,\ldots,cv_n]$, we construct $cv_i$ by first following the pointer $p_i$ and then shifting the result $h_i$ positions to the left.  Alternatively, if we can shift and add in one step, we can perform this shifting when we add $cv_i/2^{h_i}$ to the element of the result matrix to which it belongs.

\end{enumerate}

\begin{example}
\label{alignment-ex}
Suppose the given vector
$$V = [3,7,2,12,8,6]$$
When we divide out powers of two, we are left with
$$[3,7,1,3,1,3]$$
When we sort and eliminate duplicates, we have vector $S=[1,3,7]$.  The vector of pointers is $P=[2,3,1,2,1,2]$.  For instance, the first element of $V$, which is 3, appears in position 2 of $S$.  The fourth element of $V$, which is 12, has also become 3 after removing factors of 2, so the fourth element of $P$ is also 2.  The vector $H$ that records the number of positions shifted is $H=[0,0,1,2,3,1]$.  For example, the 3 in $V$ is not shifted at all, while the 12 in $V$ has been shifted two positions to the right.
\end{example}

There are two advantages to this alignment step.  First, it evidently reduces the number of elements of $V$ that are considered distinct.  Thus, it reduces the length of the sorted list and the length of the vector of differences.  But it has another, more subtle, effect.  The elements of the sorted list are all odd.  Therefore, all differences other than (perhaps) the first are even.  Thus, when called recursively the first time, differences have at most $b-1$ bits after shifting right to eliminate trailing 0's.

\subsection{Modified Scalar Multiplication}
\label{column-simp-subsect}

We have described the algorithm as taking a column of the first matrix, and processing each of its $n$ values independently.  However, if there are duplicates among those values, as will likely be the case, we should first eliminate duplicates.  Further, we do nothing for those values that are 0, and for values that are 1, no multiplication is needed; we can take the original vector $V$ as the product.

We can also apply the trick of Section~\ref{alignment-subsect} to the columns of the first matrix.  For instance, if 3 and 12 are both elements of a column, and we have computed $3V$, then we do not also need to compute $12V$; we can simply shift the values of the vector $3V$ two positions left.

\subsection{Parallelism}
\label{parallelism-subsect}

Ideally, as much of the circuitry on a chip should be active at any given time.  The algorithm we have described is a serial algorithm, so it will not tend to keep things active.  However, there are many opportunities to increase the parallelism in the chip.  First, note that other than the circuit to sort, which itself can be parallelized in known ways, e.g. \cite{batcher}, most of the components needed are either registers to store the various vectors needed, or adder circuits. Moreover, as described in Fig.~\ref{algoutline-fig}, all the vectors shown are handled with a single adder circuit.  As the accumulation of differences (step~4 in Section~\ref{vector-scalar-subsect}) appears serial, it is hard to parallelize this portion of the algorithm.\footnote{Strictly speaking, there is a parallel algorithm \cite{stone} for computing the accumulated sums of $n$ elements in $O(\log n)$ time, but this approach requires $O(n\log n)$ circuit components and takes $(n\log{n})/2$ additions, so it would negate the advantage of using adders in place of multipliers.}

We can multiply the vector $V$ by many different scalars $c$ at the same time, but we need registers to store intermediate results for each $c$.  That change may thus speed up the time needed by a large factor, but it doesn't change the ratio of space needed for registers compared with space for adders.  Likewise, we can process many different rows $V$ in parallel; that also speeds the process but doesn't change the space allocation.

There is one modification to the algorithm that {\em will} increase the ratio of adder space to register space.  After sorting and eliminating duplicates, we can break the vector $S$ into several segments: one segment for the smallest values, another for the next smallest values, and so on.  We can then process each segment independently, in parallel.  That change allows us to use many adders at once to accumulate differences for a single vector $S$.

There is a downside to this approach: When we take differences within the segments, the same difference may occur in many different segments, so we may be doing the same work several times without realizing it.  But because we create segments after sorting, the sum of all the differences among all the segments has not changed; it is still at most $2^b$.  Thus, we can still expect a significant reduction in the total length of vectors after we take second differences.  An example suggests what to expect.

\begin{example}
\label{segments-ex}
Let us reconsider the situation of Example~\ref{count-ex}, where we had 1000 12-bit integers and argued that there could be no more than 90 differences.  After sorting, we might have fewer than 1000 elements, but let us assume the sorted vector $S$ still has 1000 elements.  Suppose we divide $S$ into ten segments of 100 elements each and take differences within each segment.  The sum of all those differences is still at most 4096.

Suppose that these differences divide as evenly as possible.\footnote{It can be shown that an even distribution is the worst case; i.e., it allows the largest number of differences.}  Then each segment has differences totaling at most 410.  Since the sum of the first 29 integers exceeds 410, there can be at most 28 differences in any of the ten segments, or a total of 280 differences.  That number is much larger than the 90 differences that can occur if we do not divide $S$ into ten segments, but it is much smaller than 1000; i.e., we are still guaranteed to reduce significantly the total length of all the segments when we take differences.
\end{example}

A second possible approach is to divide a vector of length $n$ into $\sqrt{n}$ segments of length $\sqrt{n}$ each.  Accumulate the sum within each segment.  Then, accumulate the sums of the final sum of each segment, to get the value that must be added to each member of each segment.  That is, we must add to each element of the $i$th segment the sum of the last elements in each of the segments 1 through $i-1$.  This approach gives $\sqrt{n}$-fold parallelism, while requiring $2n+\sqrt{n}$ additions in place of the $n$ additions that would be needed to do a sequential accumulation of all $n$ elements.

\subsection{Extension to Other Matrix Forms}
\label{extensions-subsect}

Matrices of interest are rarely square, dense matrices of nonnegative integers.  Here is a list of some of the extensions that can be made to the basic algorithm.

\subsubsection{Positive and Negative Integers}

We have assumed that all matrix elements are nonnegative integers.  But a sign can be carried along with each element and not used until it is time to place a sign on each scalar product.  The algorithm for choosing the correct sign should be obvious.

\subsubsection{Floating-Point Numbers}

Multiplication of floating-point numbers involves multiplying the mantissas and adding the exponents.  Multiplication of the mantissas is an integer multiplication.  Adding of exponents is, of course, an addition, one that is necessary no matter how we perform the multiplication.

\subsubsection{Nonsquare Matrices}

Suppose we need to multiply a matrix $A$ that is $n$ rows by $k$ columns times a matrix $B$ that is $k$ rows by $m$ columns.  Then we need to take $k$ outer products, each of a column of $A$ times the corresponding row of $B$.  Nothing about the algorithm described so far requires that $n=m$ or that either equals $k$.  The only extension is that we need to decide whether to make the columns of $A$ or the rows of $B$ play the role of the vector $V$.  Intuitively, the longer the vector, the fewer additions per element we need.  Thus, the choice normally depends on picking columns of $A$ if $n>m$ and picking rows of $B$ otherwise.  The only time that choice might not be better is if there is so much parallelism available that we can process more than $\min(n,m)$ scalars at once.

\subsubsection{Sparse Matrices}

There are two cases to consider here.  If the matrices are represented in the ordinary manner, as arrays, then the first step, where we sort and eliminate duplicates, essentially eliminates all the 0's.  We cannot do better, because we have to look at the entire matrices, regardless of how sparse they are.

In the second case, the matrices $A$ and $B$ are represented as a set of triples $(i,j,v)$ meaning that the element in row $i$ and column $j$ has the nonzero value $v$.  We can assemble columns of $A$ or rows of $B$ by finding all the triples that belong to that row or column.  The rows and columns are missing the 0 elements, but because we have location data for each element,  we can then take the outer product of column $k$ of $A$ and row $k$ of $B$ by working with only the nonzero elements.  Any product where one or both of the arguments is 0 would yield a 0 product and thus never influence the result anyway.  The only significant modification to the algorithm is that along with the vector $V$ we need to have a parallel vector that gives, for each element, the row (if $V$ is a column of $A$) or the column (if $V$ is a row of $B$).

\section{Experimental Results}
\label{exper-sect}

Figure~\ref{results-fig} shows the lengths of the lists that result from the following experiments.  For four values of $n$ ranging from one thousand to one million, we generated 100 lists of random 24-bit numbers.  These lists were sorted and duplicates were eliminated.  In some cases, we right-shifted (aligned) the numbers first to eliminate factors of 2.  The lengths of the resulting sorted lists are shown in the column labeled (A); it and the following columns are averages rounded to the nearest integer.

\begin{figure}[ht]

\begin{center}
\begin{tabular}{|r|c|r|r|r|r|r|}
\hline
$n$ & align? & (A) & (B) & (C) & (D) & $+/*$\\
\hline\hline
$10^3$ & no & 1000 & 985 & 228 & 39 & 2.68\\
$10^3$ & yes & 1000 & 871 & 73 & 13 & 2.12\\
\hline
$10^4$ & no & 9997 & 3963 & 72 & 17 & 1.42\\
$10^4$ & yes & 9991 & 1395 & 28 & 6 & 1.15\\
\hline
$10^5$ & no & 99706 & 1170 & 22 & 7 & 1.01\\
$10^5$ & yes & 99119 & 470 & 9 & 3 & 1.00\\
\hline
$10^6$ & no & 970772 & 193 & 6 & 3 & 0.97\\
$10^6$ & yes & 917540 & 85 & 3 & 1 & 0.92\\
\hline
\end{tabular}
\end{center}

\caption{Result of sorting and taking differences three times}
\label{results-fig}

\end{figure}

Column~(B) shows the lengths of the lists after taking differences and performing the same operations on the list of differences~-- align (if permitted), sort, and eliminate duplicates.  Then, columns (C) and (D) represent the lengths of the lists that result after repeating this operation twice more.  The last column gives the average number of additions that would be needed to multiply the initial vector of length $n$ by a scalar.  To be precise, it is 12 times column~(D) (for the Russian-peasants multiplication of each element on the list of third differences), plus columns (A), (B), and (C), all divided by $n$.

\subsection{Intuition}
\label{intuition-subsect}

If we look at the first row of Fig.~\ref{results-fig}, we see a typical situation where the length of the vector, $n$, is much smaller than the number of possible integers.  In this case, the chances that a random list of 1000 integers, each chosen from around 16 million 24-bit integers, would have even one duplicate is small.  That remains true, even if we use alignment, which in effect divides integers into about 8 million groups, each corresponding to a 24-bit odd integer.  Moreover, as we see from column~(B), even the differences between elements of the sorted list are almost all distinct.  It is not until we take differences of the differences that we begin to see duplicates, as suggested by column~(C).  By the time we take third differences, there are very few distinct numbers indeed, as seen in column~(D).

In contrast, let us look at the last two rows, where there are a million random integers initially, again chosen from the roughly 16 million possible 24-bit integers.  Now there are good chances of seeing some duplication, and indeed we do.  But the most significant effect is seen in column~(B), where the number of distinct differences is tiny, even if we do not align.  The reason is that when we have a million numbers out of 16 million possibilities, the average difference in the sorted list is only 16.  Surely, there will be some larger gaps, but the chances of a really large gap, say over 100, is small, and there cannot be too many of those large gaps, because the sum of all the gaps is at most $2^{24}$.  As a result, the total work dealing with all the difference lists is negligible, and the vast majority of the work occurs computing the result of multiplying the list represented by column~(A).  In fact, we require less than one addition per scalar multiplication, because the number of duplicates in the original list exceeds the length of all the difference lists.

The effect of alignment is perhaps more profound than might be expected, especially for the smaller values of $n$.  Our first assumption might be that alignment doubles the number of duplicates.  That is, the 16 million possible integers are grouped into 8 million groups, each corresponding to an odd number.  That would appear to double the chances that a randomly chosen integer has appeared before on the initial list.  But we are in fact beneficiaries of the ``class-size paradox.''  The groups are not all the same size.  For example, the group to which the integer 3 belongs has 23 members~-- all the 24-bit integers that are all 0's except for two conscutive 1's somewhere.  So a random integer tends to belong to a larger group, and further it will be a duplicate if any of the members of that group have appeared previously.  For example, if we compare the last two lines, we see that if we do not align the one million integers, we find an average of 29,228 duplicates, while if we {\em do} align we average 82,460 duplicates, almost three times as many.

\section{An Upper Bound on Required Addition Steps}
\label{upper-sect}

The results of Section~\ref{exper-sect} are based on the assumption that vectors are chosen randomly and uniformly.  Especially for long vectors, randomness forces difference lists to be small.  However, in many applications it is not reasonable to assume uniformity in the distribution of values; there might well be many small values but also enough large values that the differences in the sorted list are mostly distinct.

We shall show that, regardless of the initial list, computing a scalar-vector product using only additions and the operations needed to sort, eliminate duplicates, align bit strings, and follow pointers, but not multiplication, requires only a small number of additions per element of the vector.  The exact number of additions per element depends on the relationship between the length of the vector and the size of its elements, but it is certainly much less than the baseline of Section~\ref{rpm-subsect} (except perhaps in the uninteresting case of very small matrices).

Here is the intuition behind the proof.  As we mentioned, when we have a list of integers, each no more than $k$, and we sort, eliminate duplicates, and take differences, the sum of those differences is at most $k$.  The list of differences may have some small numbers, say numbers at most $x$~-- a value we shall select judiciously.  But when we take differences a second time, there cannot be many differences among the small numbers, since their sum is at most $x$.  There may also be some larger numbers, those bigger than $x$.  However, there can be at most $k/x$ of those, since the sum of all the numbers is at most $k$.  Thus, when taking differences a second time, there cannot be many differences among the large numbers either.  These two observations let us put a strong upper bound on the lengths of the lists as we repeat the sort-eliminate duplicates-take differences process.

\subsection{Two Cost Functions}
\label{CD-subsect}

It is useful to define two mutually recursive cost functions, $C$ and $D$.

\begin{itemize}

\item
Define $C(n,k)$ to be the number of additions needed to multiply, by a constant, a vector of $n$ elements, each of which is a positive, odd integer no larger than $k$, by a constant.\footnote{Note that $k$ is $2^b$ if we are talking about $b$-bit integers.}

\item
Define $D(n,k)$ to be the number of additions needed to multiply, by a constant, a vector of length $n$, whose elements are distinct, odd positive numbers that sum to $k$.

\end{itemize}
Observe that the significant difference between $C$ and $D$ is that in the former case, $k$ is a bound on each individual element, while in the latter case, $k$ bounds the sum of all the elements, and therefore represents a more limited class of vectors.

\subsection{Bounds on the Cost Functions}
\label{CD-bounds-subsect}

We can observe four rules that let us put bounds on $C(n,k)$ and $D(n,k)$ mutually.

\medskip

{\bf Rule 1}: $C(n,k) \le D(n,k) + n$.  This rule reflects the idea that we can align the given vector $[v_1,v_2,\ldots,v_n]$ of length $n$, sort, eliminate duplicates, and take differences of the resulting vector.  We shall surely have a difference vector of length no greater than $n$, and the sum of all its elements cannot exceed $k$, the maximum possible element in the original vector.  Recursively multiply this vector by a constant $c$.  We can then compute the value of each $cv_i$ by accumulating the differences.  This step uses at most $n$ shift-and-add additions.  The shifting is necessary because some of the differences have had factors of 2 removed during the alignment process.  The only other additions needed are in the recursive computation of the product of $c$ and the vector of differences.  That cost is no greater than $D(n,k)$.

\medskip

{\bf Rule 2}: $D(n,k) \le C(n,x) + D(k/x,k)$ for any $x$.  This rule doesn't involve any operations, per se.  Rather, it states that we can conceptually treat the vector of length $n$ as two vectors.  The first consists of the ``front'' part~-- those elements less than or equal to $x$.  The second is the ``back'' part~-- elements greater than $x$.   For the front, we have an upper bound on each element; they can be no larger than $x$.  We also have the condition from the definition of $D$, which is that the sum of these elements can be no greater than $k$, but we shall not use that. Presumably, the upper bound $x$ on each element will prove stronger.  For the back part, we know that the sum of these elements is not greater than $k$.  We also know that there can be no more than $k/x$ of these elements, since each is greater than $x$.

\medskip

{\bf Rule 3}: $D(n,k) \le D(\sqrt{k},k)$.  Each of the elements on a list is odd and the elements are distinct.  Therefore, given that numbers sum to $k$, the longest the list can be is that $r$ such that $1+3+5+\cdots+2r-1 \le k$.  Since the sum of the first $r$ odd numbers is exactly $r^2$, it follows that $\sqrt{k}$ is actually an upper bound on the length of the list described by $D(n,k)$.  Of course this observation is only valuable if $n>\sqrt{k}$, but the inequality of the rule is true regardless.

\medskip

{\bf Rule 4}: $C(n,k)$ and $D(n,k)$ are each no larger than $n\log_2k$.  This rule holds because we could always use the Russian-peasants algorithm if nothing better is available.

\medskip

We can use these rules to prove the following theorem.  It says, roughly, that the number of additions per element you need to multiply a vector of length $n$ by a constant, with elements no larger than $k$, is $\log_nk$.

\begin{theorem}
\label{main-th}
If
$$n \ge \Bigl(\frac{j+1}{2}\Bigr)k^{1/j}\log_2k$$
then $C(n,k) \le jn$.
\end{theorem}

\begin{proof}
Start with $C(n,k)$, but let us replace $k$ by $x_0$ for reasons that will become obvious.  By Rule~1 we have
$$C(n,x_0) \le D(n,x_0) + n$$
Apply Rule~2 to give us
$$C(n,x_0) \le C(n,x_1) + D(\frac{x_0}{x_1},x_0) + n$$
We may alternate Rules 1 and 2 as many times as we like, ending with Rule~1, and introducing a new unknown $x_j$ each time we do, to get
$$C(n,x_0) \le D(n,x_i) + \sum_{j=0}^{i-1} D(\frac{x_j}{x_{j+1}} ,x_j) + (i+1)n$$
Next, apply Rule~3 to the first term on the right, which gives us
$$C(n,x_0) \le D(\sqrt{x_i}, x_i) + \sum_{j=0}^{i-1} D(\frac{x_j}{x_{j+1}} ,x_j) + (i+1)n$$
Now, we choose values for each of the $x_j$'s in order to make the terms equal.  That is, pick $x_p = k^{(i+2-p)/(i+2)}$ for $p=0,1,\ldots,i$.  In particular, $x_i = k^{2/(i+2)}$, so the first term $D(\sqrt{x_i},x_i)$ becomes $D(k^{1/(i+2)},k^{2/(i+2)})$.  The summation $\sum_{j=0}^{i-1} D(\frac{x_j}{x_{j+1}} ,x_j)$ becomes $\sum_{j=0}^{i-1} D(k^{1/(i+2)}, k^{(i+2-j)/(i+2)})$.  The first term $D(k^{1/(i+2)},k^{2/(i+2)})$ fits into this sum so we can write
$$C(n,k) \le \sum_{j=0}^i D(k^{1/(i+2)}, k^{(i+2-j)/(i+2)}) + (i+1)n$$
Note that on the left, we replaced $x_0$ by the original value $k$, for which it stood.

Finally, we use Rule~4 to bound each of the terms in the above sum.  That gives
$$C(n,k) \le k^{1/(i+2)} \log_2k \sum_{j=0}^i \frac{i+2-j}{i+2} + (i+1)n$$
The summation is $1/(i+2)$ times $2+3+\cdots+(i+2)$.  The latter is the sum of the first $i+2$ integers, although it is missing 1.  That sum is therefore $\frac{(i+2)(i+3)}{2} -1$.  We may drop the ``$-1$'' and conclude
$$C(n,k) \le \frac{i+3}{2}k^{1/(i+2)}\log_2k + (i+1)n$$
As long as the first term is at most $n$, we have $C(n,k) \le (i+2)n$.
To simplify, substitute $j$ for $i+2$.  We can then assert that if $n \ge \frac{j+1}{2}k^{1/j}\log_2k$ then $C(n,k) \le jn$.
\end{proof}

\begin{example}
\label{proof-ex}
Suppose $k=2^{24}$.  Then $2n$ additions suffice if $n \ge \frac{3}{2}\sqrt{k}\log_2k = 147{,}456$.  Also, $3n$ additions suffice if $n \ge 2k^{1 / 3} \log_2k = 12{,}288$.  One needs at most $4n$ additions if $n \ge \frac{5}{2} k^{1/4 } \log_2k = 3840$.
\end{example}

\begin{example}
\label{fib-ex}
The purpose of this example is to address concerns one may have regarding extreme cases where the process sort-elim\-inate-dupli\-cates-take-differ\-ences makes lit\-tle progress.  Such cases exist, but they require that the size of the matrix be very small~-- comparable to the number of bits used to represent elements.

For instance, suppose we start with powers of 2.  If we do not align, then when we take differences we get the original vector with the highest power of 2 removed.  However, in this case, $n$ is equal to $\log_2k+1$, and the hypothesis of Theorem~\ref{main-th} cannot be satisfied except in trivial cases~-- specifically, for $k=1$ (i,e,, a Boolean matrix), and the case $k=2$, $j=1$..  Moreover, if we align, then all powers of 2 immediately become 1, and we actually need no additions at all, just shifts.

Another seemingly bad case is the Fibonacci numbers, where after sorting and taking differences we only lose the two highest values.  For example, starting with the vector 
$$[1,2,3,5,8,13,21]$$
when we take differences we get $[1,2,3,5,8]$.  Here, alignment doesn't help.  But we still have the constraint that $n$ must be logarithmic in $k$, although the base of the logarithm is now 1.61, approximately.  It is still not possible to find nontrivial values for $n$, $k$, and $j$ to satisfy the hypothesis of Theorem~\ref{main-th}.
\end{example}

\section{Attribution}

Note: The algorithm described here was invented solely by Daniel Cussen.  The proof of Section~\ref{upper-sect} is solely the work of Jeffrey Ullman.

\bibliographystyle{abbrv}
\bibliography{references}